\documentstyle[12pt]{article}
\begin{document}

\begin{titlepage}
\begin{flushright}
DPSU-95-11\\
October 1995
\end{flushright}

\vskip 2cm

\begin{center}
{\Large On Low-Energy Theory from General Supergravity\footnote{
Talk presented at YKIS'95, Kyoto, Japan, August, 1995}}

\vskip 2cm

{\Large Yoshiharu~Kawamura\footnote{E-mail address:
ykawamu@gipac.shinshu-u.ac.jp}}

\vskip 0.5cm

{\large\sl Department of Physics, Shinshu University\\
Matsumoto, 390 Japan}
\end{center}

\vskip 2cm

\begin{center}
{\bf ABSTRACT}
\end{center}
Starting from non-minimal supergravity theory
with unified gauge symmetry,
we obtain the low-energy effective theory
by taking the flat limit and integrating out the
superheavy fields in a model-independent manner.
The scalar potential has extra non-universal contributions
to soft supersymmetry breaking terms
which can give an impact on phenomenological study.

\end{titlepage}

\newpage

\section{Introduction}

The standard model (SM) has been
established as an effective theory below the weak scale,
although at present there are some measurements inconsistent
with the SM predictions.\cite{Hagiwara}
The search for the theory beyond SM is one of the most
important subjects in elementary particle physics.
SM has a problem called
$\lq$naturalness problem'.\cite{naturalness}
This problem essentially means that there is no natural
mechanism to keep the value of Higgs field's mass
the weak scale one against radiative corrections, and it
can be a key to explore new physics.
In fact, $\lq$naturalness problem' is elegantly solved by the
introduction of $\lq$supersymmetry' (SUSY).\cite{SUSY}

The minimal SUSY extension of SM (MSSM) is regarded
as a candidate of realistic theory beyond SM.\cite{MSSM}
The Lagrangian density of MSSM consists of two parts,
\begin{eqnarray}
{\cal L}_{MSSM} &=& {\cal L}_{MSSM}^{SUSY} + {\cal L}_{MSSM}^{Soft},
\label{L}\\
{\cal L}_{MSSM}^{Soft} &=& -{1 \over 2}\sum_a M_a
\lambda^a \lambda^a - H.c.
- \sum_{k,l} (m^2)_k^l z^k z^*_l \nonumber\\
&~& - \sum_{k,l,m} A_{klm} z^k z^l z^m
    - \sum_{k,l} B_{kl} z^k z^l - H.c. ,
\label{Lsoft}
\end{eqnarray}
where ${\cal L}_{MSSM}^{SUSY}$ is the SUSY part and
${\cal L}_{MSSM}^{Soft}$ is the soft SUSY breaking part.
Here $\lambda^a$'s $(a=1,2,3)$ are gauginos
(bino, wino, gluino) and $z^k$'s are scalars
(squarks, sleptons and Higgs doublets).
%Soft Susy breaking terms consist of gaugino mass terms,
%scalar mass terms, scalar trilinear
%coupling terms and bilinear coupling terms.
The parameters $(M_a, (m^2)_k^l, A_{klm}, B_{kl})$
are called $\lq$soft SUSY
breaking parameters' and they are arbitrary and the origin
is unknown in the MSSM.\footnote{
In this paper, we do not assume the universality on the
soft SUSY breaking parameters from the beginning
 when we use the terminology $\lq$MSSM'.}

It is expected that these parameters originate in more
fundamental theories.
We have quite an interesting scenario for the origin of
soft SUSY breaking terms based on supergravity (SUGRA).\cite{SUGRA}
The SUSY is spontaneously or dynamically broken in the so-called
hidden sector and the effect is transported to our observable
sector by the gravitational interaction.
As a result, soft SUSY breaking terms appear in our sector.
In this scenario, the pattern of soft SUSY breaking terms is
determined by the structure of SUGRA.
For example, it is well-known that the minimal SUGRA leads to
a universal type of soft SUSY breaking parameters.
The scalar potential $V$ is given as follows,\cite{Hidden}
\begin{eqnarray}
V &=& V_{SUSY} + V_{Soft},
\label{Min-V}\\
V_{SUSY} &=& |\frac{\partial \widehat{W}}{\partial z^k}|^2
      + {1 \over 2}g_a^2
(z_k^* (T^a)^k_l z^l)^2 ,
\label{Min-VSUSY}\\
V_{Soft} &=& A \widehat{W}
+ B z^k \frac{\partial \widehat{W}}{\partial z^k}
		+ {\it H.c.}
	+ |B|^2 z_k^* z^k,
\label{Min-Vsoft}
\end{eqnarray}
where $\widehat{W}$ is a superpotential,
$g_a$'s are gauge coupling constants
and $T^a$'s are gauge generators.
$V_{SUSY}$ stands for the SUSY part, while $V_{Soft}$
 contains the soft SUSY breaking terms.
The parameters $A$ and $B$
 are written as
\begin{eqnarray}
A &=& {\langle \tilde{F}^i \rangle
\langle \tilde{z}^{\ast}_i \rangle \over M^2}
 - 3m_{3/2}^{\ast},
\label{A}\\
B &=& m_{3/2}^{\ast},
\label{B}
\end{eqnarray}
where $\tilde{F}^i$'s and $\tilde{z}^i$'s are $F$-components
and scalar components of chiral supermultiplets
in the hidden sector, respectively.
The bracket $\langle \cdots \rangle$ denotes the vacuum
expecectation value (VEV) of the quantity,
$M$ is a gravitational scale and $m_{3/2}$ is a gravitino mass.

On the other hand, SUSY-Grand Unified Theory (SUSY-GUT)
\cite{SUSY-GUT} has been
hopeful as a realistic theory.
In fact, the precision measurements at LEP\cite{LEP}
 have shown that the gauge coupling constants
$g_3$, $g_2$ and $g_1$ of $\lq$SM gauge group'
$G_{SM} = SU(3)_C \times SU(2)_L \times U(1)_Y$
meet at about
 $10^{16}$ GeV
 within the framework of MSSM.\cite{LAM}
SUSY $SU(5)$ GUT is the simplest unification scenario
 and predicts the long lifetime of nucleon
consistent with the present data.\cite{HMY}
However various unification scenarios
consistent with the LEP data
 have been known within SUSY-GUTs.
For example, the direct breaking of the larger group
 down to $G_{SM}$ and
 the models with extra heavy generations.
Non-trivial examples are
the models of SUSY $SO(10)$ GUT with chain breaking.
\cite{Deshpande,KMY1,BST}
So it is important to specify the realistic SUSY-GUT model by
using some observables in addition to gauge couplings.

Here let us emphasize that the soft SUSY breaking parameters can be
powerful probes for physics beyond the MSSM such as SUSY-GUTs,
SUGRAs, and superstring theories (SSTs).
The reason is as follows.
The SUSY spectrum at the weak scale, which is expected to
be measured in the near future, is translated into the soft
SUSY breaking parameters.
And the values of these parameters at higher energy scales
are obtained by using the renormalization
group equations (RGEs).\cite{RGE}
In many cases, there exist, at some energy scale,
some relations among these parameters.
They reflect the structure of high-energy physics.
Hence we can specify the high-energy physics by checking
these relations.

We give some examples.\footnote{
We neglect the threshold corrections, the effect of higher
dimensional operators,
the mass mixing effect and so on.}
\begin{enumerate}
\item We can know whether the $\lq$SM gauge group' is grand-unified
or not by checking the $\lq$GUT relation' among gaugino masses
$M_a, (a=1,2,3)$
\begin{eqnarray}
  \frac{M_1}{5/3 g^2_Y}=\frac{M_2}{g_2^2}
 =\frac{M_3}{g_3^2}.
 \label{gaugino}
\end{eqnarray}
It is shown that the gaugino mass spectrum satisfies
the $\lq$GUT relation' at any energy scale between the
unification scale and the weak scale
as far as the $\lq$SM model gauge group' is embedded into
a simple group, irrespective of
the symmetry breaking pattern.\cite{KMY1}

\item The pattern of gauge symmetry breakdown can be specified
by checking certain sum rules among scalar masses.
For example, the scalar masses satisfy the following
mass relations for the breaking $SU(5) \to G_{SM}$
\begin{eqnarray}
&~&m_{\tilde{q}}^2 = m_{\tilde{u}}^2 = m_{\tilde{e}}^2
 \equiv m_{10}^2 ,\\
&~&m_{\tilde{l}}^2 = m_{\tilde{d}}^2 \equiv m_{5^*}^2 ,
\label{scalar}
\end{eqnarray}
at the breaking scale.
Here $m_{\tilde{q}}$, $m_{\tilde{u}}$, ... are soft SUSY breaking
scalar masses of squark doublet $\tilde{q}$,
up-type singlet squark $\tilde{u}$ and so on.
Scalar mass relations are derived
for $SO(10)$ breakings\cite{KMY1} and
for $E_6$ breakings.\cite{KT}

\item We can know the structure of SUGRA and SST by checking some
specific relations among soft SUSY breaking parameters.
For example, the SST with the SUSY breaking due to dilaton
$F$-term leads to the highly restricted pattern
such as\cite{dilaton-F}
\begin{eqnarray}
-A = M_{1/2} = \sqrt{3} m_{3/2}
\label{dilaton}
\end{eqnarray}
where gauginos and scalars get masses
with common values $M_{1/2}$ and $m_{3/2}$, respectively.
\end{enumerate}

In this way, the soft SUSY breaking parameters can play important
roles to probe new physics, but here we should note that the
features of these parameters have not been completely
investigated based on SUGRA with a general structure yet.

We have two important consequences so far.

(1) The precision measurements of the SUSY spectrum are very
important.
We hope that projects using next-generation colliders
are developed and advanced quickly.

(2) But first it is important to place the low-energy theory
within a more general framework as it relates to SUGRA.
This is the motivation of our work.\cite{K}

The content of this paper is as follows.
In section 2, we briefly show the procedure of
the derivation and the result of our low-energy
theory.
We give a conclusion in section 3.

\section{The derivation and the result}

Various types of low-energy theories have been derived
based on the hidden sector SUSY breaking scenario in a
model-dependent or model-independent way.
\cite{Hidden}\cite{HLW,SW,D,KMY2}
The difference among their structures arises from what type of
SUGRA has been taken as a starting point.
Four types of SUGRAs occur to us, that is,
the minimal one, the minimal one with GUT,
non-minimal one and non-minimal one
with GUT.
The first three cases have been energetically
investigated.\cite{Hidden}\cite{HLW}\cite{SW}
The study of the last case has also been started in a
model-independent manner.\cite{KMY2}

Let us explain the work of Ref.\cite{KMY2} briefly.
The starting theory is a SUSY-GUT with non-universal soft SUSY
breaking terms, which is derived from non-minimal SUGRA
with a {\it hidden} ansatz by taking the flat limit first.
Here the {\it hidden} ansatz means that the superpotential is
separate from hidden sector to the observable one such as
%We take the {\it hidden} ansatz for the superpotential as
$W_{SG} = W(z) + \tilde{W}(\tilde{z})$.
It is shown that there exist extra non-universal contributions
 to soft SUSY breaking terms and
some phenomenological implications are discussed.
The results are written down in terms of SUSY-GUT,
so it might be relatively easy to compare the values of
measurements with the parameters in SUSY-GUT in the future.
But we could have wished to know the information on
the structure of SUGRA directly.
Hence we would like to carry out the following subjects
(1) to take a more general SUGRA,
e.g. to take off the {\it hidden} ansatz
(2) to write down the low-energy theory in terms of SUGRA
in order to connect the experiments with SUGRA directly.

Our setting is SUGRA with
non-minimal K\"ahler potential and a certain unified gauge
symmetry.
And our goal is to obtain its low-energy theory by taking
the flat limit and integrating out heavy fields
in a model-independent way.

First we give some basic assumptions.
\begin{enumerate}
\item The SUSY is spontaneously broken by the $F$-term
condensation in the hidden sector.
The Planck scale physics plays an essential role
in the SUSY breaking.
The hidden fields $\tilde{z}^i$ are gauge singlets
and they have the VEVs of $O(M)$.
The magnitude of ${W}_{SG}$ and $\tilde{F}^i$
are $O(m_{3/2} M^2)$ and $O(m_{3/2} M)$, respectively.
We identify the gravitino mass with the weak scale.\footnote{
This assumption may be a little too strong since we only need to
require that the soft SUSY breaking masses are of order of
weak scale.
In fact, there is quite an interesting scenario\cite{GLM}
that the gravitino mass is decoupled to the soft parameters
and the magnitude of the SUSY breaking is determined by the
gaugino masses.}

\item The unified gauge symmetry is broken down at
the unification scale $M_U$ independent of the SUSY breaking.
Some observable scalar fields have the VEVs of $O(M_U)$.

\item All fields are classified into two categories
by using the values of those masses.
One is a set of heavy fields with mass of $O(M_U)$.
The other is a set of light fields with mass of $O(m_{3/2})$.
There are no light singlet observable fields
which induce a large tadpole contribution
 to Higgs masses by coupling to
Higgs doublets renormalizably in superpotential.
\end{enumerate}

Next we explain the procedure to obtain the low-energy
theory.
\begin{enumerate}
\item We calculate the VEVs
 of derivatives and write down the scalar potential by using
the flactuations $\Delta z$.

\item When there exists a mass mixing, we need to diagonalize
the scalar mass matrix to identify the
heavy fields and the light ones correctly.

\item Then we solve the stationary conditions of the potential
for the heavy fields while keeping the light fields arbitrary
and integrate out
the heavy fields by inserting the solutions into
the scalar potential.
\end{enumerate}

On the derivation of the scalar potential,
we come across a problem related to the stability of
the weak scale.
The problem is as follows.
Some light fields, which contain weak Higgs doublets,
classified by using SUSY fermionic masses generally
would get intermediate masses at tree level
after the SUSY is broken down.
We explain it by taking SUGRA without a unified symmetry
as an example.
When the {\it hidden} ansatz is taken off,
the following extra terms should be added,
\begin{eqnarray}
\frac{\partial \widehat{W^*}}{\partial \tilde{z}^*_i}
\langle (K^{-1})_i^j \rangle
\frac{\partial \widehat{W}}{\partial \tilde{z}^j}
 + \Delta C(z, z^*)
+ \langle \tilde{F}^i \rangle
\frac{\partial \widehat{W}}{\partial \tilde{z}^i}
		+ {\it H.c.},
\label{ExtraV}
\end{eqnarray}
where $\Delta C(z, z^*)$ is a bilinear polynomial of $z$ and $z^*$.
The magnitude of the third term and its hermitian conjugate
can be of order $m_{3/2}^3 M$
if the Yukawa couplings between the hidden sector fields
and the observable sector light fields are of order unity,
and so a large mixing mass of
Higgs doublets can be introduced.
In the presence of such a large $B$-parameter, the electro-weak
symmetry breaking does not work at the weak scale.
Hence we require that such dangerous terms are suppressed as
\begin{eqnarray}
 \langle \tilde{F}^i \rangle
\frac{\partial \widehat{W}}{\partial \tilde{z}^i}
= O(m_{3/2}^4) ,
\label{gh}
\end{eqnarray}
by some mechanism.
This requirement gives a constraint on the total K\"ahler
potential.
Of course, models with the hidden ansatz fulfill
this requirement trivially.
In the same way, we must impose some conditions to keep
the gauge hierarchy in the case of SUGRA with unified
gauge symmetry.\cite{JKY}\cite{K}

Our SUGRA consists of the K\"ahler potential $K$,
the superpotential
$W_{SG}$ and the gauge kinetic function $f_{\alpha\beta}$,
which are written down in terms of
the variations $\Delta \hat{z}^{\hat{I}}$ of mass eigenstates
as follows,
\begin{eqnarray}
K
&=&\langle \hat{K} \rangle + \langle \hat{K}_{\hat{I}} \rangle
 \Delta \hat{z}^{\hat{I}}  + {1 \over 2}\langle
\hat{K}_{\hat{I}\hat{J}} \rangle
\Delta \hat{z}^{\hat{I}} \Delta \hat{z}^{\hat{J}}
+ \cdots ,
	\label{hatK}\\
W_{SG}
&=&\langle \hat{W} \rangle + \langle \hat{W}_{\hat{I}} \rangle
 \Delta \hat{z}^{\hat{I}}  + {1 \over 2}\langle
\hat{W}_{\hat{I}\hat{J}} \rangle
\Delta \hat{z}^{\hat{I}} \Delta \hat{z}^{\hat{J}}
\nonumber \\
&~&   + {1 \over 3!}\langle \hat{W}_{\hat{I}\hat{J}\hat{J'}}
\rangle \Delta \hat{z}^{\hat{I}} \Delta \hat{z}^{\hat{J}}
\Delta \hat{z}^{\hat{J'}}
+ \cdots
	\label{hatW}
\end{eqnarray}
and
\begin{eqnarray}
f_{\alpha\beta} &=& f_{\alpha\beta}(\Delta \hat{z}) ,
\label{f}
\end{eqnarray}
where $\hat{I} = (I, \bar{I})$ and the ellipses represent
higher order terms.
Here we shall explain our notations for the field's indices.
The index $I$, $J$, ... run all scalar species.
In them, $i$, $j$,... and $\kappa$, $\lambda$,... run
the hidden fields and the observable ones, respectively.
Furthermore, in the observable sector fields,
$k$, $l$,..., $K$, $L$,... and $A$, $B$,...
run the light non-singlet fields, the heavy complex ones
and the heavy real ones related to the broken generators,
respectively.

Under the above-mensioned assumptions and requirements,
we can obtain the scalar potential ${V}^{eff}$
by the straightforward calculation.
The result can be compactly expressed if we define the effective
superpotential $\widehat{W}_{eff}$ as
\begin{eqnarray}
    \widehat{W}_{eff} (z) &=&
  {1 \over 2!}\hat{\mu}_{kl} \delta \hat{z}^k \delta \hat{z}^l
+  {1 \over 3!}\hat{h}_{klm} \delta \hat{z}^k \delta \hat{z}^l
\delta \hat{z}^m ,
\label{calWeff}
\end{eqnarray}
where
\begin{eqnarray}
  \hat{\mu}_{kl} &\equiv& E^{1/2}\biggl(\langle
\hat{W}_{kl} \rangle + {\langle \hat{W} \rangle \over M^2}
\langle \hat{K}_{kl} \rangle
 - \langle \hat{K}_{kl\bar{i}} \rangle \langle
(\hat{K}^{-1})^{{\bar{i}}j} \rangle
 \delta \hat{\cal G}_j \biggr)
%\nonumber \\
%&~&~~~~
+ (m^{'''}_{3/2})_{kl} ,
\label{hat-mu}\\
 \hat{h}_{klm} &\equiv& E^{1/2} \langle \hat{W}_{klm} \rangle.
\label{hat-h}
\end{eqnarray}
Here $E \equiv \langle exp(K/M^2) \rangle$,
$\langle (\hat{K}^{-1})^{{\bar{i}}j} \rangle$ is
the inverse matrix of
$\langle \hat{K}_{{\bar{i}}j} \rangle$
and $\delta {\hat{\cal G}_j} =  \langle \hat{W}_j \rangle
+\langle \hat{W} \rangle \langle \hat{K}_{j} \rangle /M^2$.
Then we can write down the scalar potential $V^{eff}$
as\footnote{
Here we omitted the terms irrelevant to the gauge non-singlet
fields $\delta \hat{z}^{\hat{k}}$ and the terms whose magnitudes
are less than $O(m_{3/2}^4)$.}
\begin{eqnarray}
{V}^{eff}&=& {V}_{SUSY}^{eff} + {V}_{Soft}^{eff},
\label{calVeff}
\\
{V}_{SUSY}^{eff} &=&
|\frac{\partial \widehat{W}_{eff}}{\partial \hat{z}^k}|^2
+ {1 \over 2}g_a^2 (\hat{z}^{\bar{k}} (T^a)_{\bar{k}l}
 \hat{z}^l)^2,
\label{calVeffSUSY}
\\
{V}_{Soft}^{eff}&=& A \widehat{W}_{eff}
   + B^k(\hat{z})_{eff}
\frac{\partial \widehat{W}_{eff}}{\partial \hat{z}^k}
		+ {\it H.c.}
\nonumber \\
&~&+ B^k(\hat{z})_{eff}{B_k(\hat{z})}_{eff} + C(\hat{z})_{eff}
+ \Delta {V},
\label{calVeffsoft}
\end{eqnarray}
where $\Delta {V}$
 is a sum of contributions from a unified symmetry breaking
and a mass mixing.
The parameters $A$, $B^k(z)_{eff}$
 and $C(\hat{z})_{eff}$ are given as
\begin{eqnarray}
A &=& m_{3/2}^{\ast '} - 3m_{3/2}^{\ast},
\label{Aagain}\\
B^k(\hat{z})_{eff} &=& (m_{3/2}^{\ast}
+ m_{3/2}^{\ast ''}+m_{3/2}^{\ast '''})_{{\bar{k}}l}
\delta^{\bar{k}k} \delta \hat{z}^l
\label{Bkeff}
\end{eqnarray}
and
\begin{eqnarray}
C(\hat{z})_{eff} &=&
E \delta \hat{\cal G}_{\bar{i}} \langle
(\hat{K}^{-1})^{\bar{i}j} \rangle
\biggl({1 \over 3!}\langle \hat{W}_{jIJJ'} \rangle \delta \hat{z}^I
\delta \hat{z}^J \delta \hat{z}^{J'}
+ {\langle \hat{W} \rangle \over M^2}
\delta^{2'}\hat{K}_j \biggr) + H.c.
\nonumber \\
&~&+ E\biggl(\delta \hat{\cal G}_{\bar{i}}
\delta^{2'}(\hat{K}^{-1})^{\bar{i}j}
\delta \hat{\cal G}_{j} + {\langle V \rangle \over M^2}
\delta^{2'} \hat{K} \biggr)
\nonumber \\
&~& - (m_{3/2}^{*'''})_{l\bar{l}} \delta^{k\bar{l}}
(m_{3/2}^{'''})_{k{\bar{k}}} \delta \hat{z}^{\bar{k}} \delta
\hat{z}^l
 - (m_{3/2}^{'''})_{kl} \delta^{k\bar{l}}
(m_{3/2}^{*'''})_{\bar{k}\bar{l}} \delta \hat{z}^{\bar{k}} \delta
\hat{z}^l
\nonumber \\
&~& - \{ (m_{3/2}^{'''})_{ml}
\delta^{m\bar{k}} (m_{3/2}^*+ m_{3/2}^{*''})_{{\bar{k}}k}
\delta \hat{z}^{k} \delta \hat{z}^{l} + H.c. \}
\nonumber \\
&~& + A \biggl[ E^{1/2}\biggl( {\langle \hat{W} \rangle \over M^2}
 \langle \hat{K}_{kl} \rangle
 - \langle \hat{K}_{kl\bar{i}} \rangle \langle
(\hat{K}^{-1})^{{\bar{i}}j} \rangle
 \delta \hat{\cal G}_j \biggr)
%\nonumber \\
%&~&~~~
+ (m_{3/2}^{'''})_{kl} \biggr] \delta \hat{z}^k \delta
\hat{z}^l ,
\label{Ceff}
\end{eqnarray}
where
\begin{eqnarray}
(m_{3/2})_{k\bar{l}} &=& E^{1/2}{\langle \hat{W} \rangle \over M^2}
 \delta_{k{\bar{l}}} ,
\label{mkl*}\\
m_{3/2}^{'} &=& E^{1/2}{\langle \hat{K}_{\bar{i}} \rangle \over M^2}
\langle (\hat{K}^{-1})^{{\bar{i}}j} \rangle \delta\hat{\cal G}_j,
\label{m'}\\
(m_{3/2}^{''})_{k\bar{l}} &=& -E^{1/2}
 \langle \hat{K}_{k\bar{l}\bar{i}} \rangle \langle
(\hat{K}^{-1})^{{\bar{i}}j} \rangle  \delta\hat{\cal G}_j,
\label{m''}\\
(m_{3/2}^{'''})_{\kappa\hat{l}} &=& -E^{1/2}
 \langle \hat{K}_{\kappa \hat{l}\bar{A}} \rangle \langle
(\hat{K}^{-1})^{{\bar{A}} \lambda} \rangle  \delta
\widehat{{\cal G}}_{\lambda}.
\label{m'''kl*}
\end{eqnarray}
Here $\widehat{\cal G}_{\lambda} \equiv
\hat{W}_{\lambda} + \hat{K}_{\lambda}\hat{W}/M^2
+ (\hat{K})_{\lambda\bar{\nu}}
 (\hat{K}^{-1})^{{\bar{\nu}}j} \hat{\cal G}_j$.
And the quantities with a prime such as $\delta^{2'} \hat{K}$
 mean that the terms proportional to
$\delta^{2} \hat{z}^{\hat{I}}$ are omitted.

There exist extra {\it chirality-conserving} scalar mass terms in
$\Delta \widehat{V}$.
The formula of the scalar masses is given as
\begin{eqnarray}
(m^2)_{k\bar{l}} &=& (m^2_0)_{k\bar{l}}
+ \langle Ref^{-1}_{AB} \rangle \langle \hat{D}^A \rangle
(T^B)_{k\bar{l}} \nonumber\\
&~& + (\mbox{$F$-term contributions}) ,
\label{mass}\\
\langle \hat{D}^A \rangle &=& 2(M_{V}^{-2})^{AB} E
 \delta \widehat{\cal G}_{\kappa} \delta \widehat{\cal G}
_{\bar{\lambda}}
 \{ G_{\bar{\mu}}^{\kappa\bar{\lambda}} (\hat{z} T^B)^{\bar{\mu}}
   + G^{\bar{\mu}\kappa} (T^B)_{\bar{\mu}}^{\bar{\lambda}} \},
\label{<hatD>}
\end{eqnarray}
where $(m^2_0)_{k\bar{l}}$'s are present before the
heavy sector is integrated out and so
they respect the original unified gauge symmetry.
And $(M_V^2)^{AB}$'s are heavy gauge boson masses and
$G=K+M^2ln(|W_{SG}|^2/M^6)$.
The most important one comes from
the $D$-term condensation of the heavy gauge sector.
It is the second term in Eq. (\ref{mass}) and
referred to as the $D$-term contribution.\footnote{
Historically, it was demonstrated that the $D$-term contribution
occurs when the gauge symmetry is broken at an intermediate scale
due to the non-universal soft scalar masses in
Refs.\cite{Hagelin} and its existence in a more general situation
was suggested in Ref.\cite{Faraggi}.}
The sizable $D$-term contribution can appear at $M$ when
the K\"ahler potential has a non-minimal structure and the rank
of gauge group is reduced by the symmetry breaking.
We can see that the $D$-term condensations
$\langle \hat{D}^A \rangle$
vanish up to $O(m_{3/2}^4/M_U^2)$ at $M$\footnote{
The $D$-term contribution can be sizable at $M_U$ by
radiative corrections even when it vanish at $M$.}
when the K\"ahler potential has the minimal structure
in the absence of Fayet-Iliopoulos $D$-term.
The $D$-term contribution is proportional to the charge of
the broken $U(1)$ factor and gives mass splittings within
the same multiplet in the full theory.
So its existence will give an impact on the
phenomenological study on the scalar masses.
\cite{KMY1}\cite{KMY2}\cite{D-term}

%We can show that our scalar potential $V^{eff}$ reduces to
%that in Ref.\cite{KMY2} after we impose an ansatz of $G$
%and take some limit.

The scalar potential obtained should be regarded as the effective
theory renormalized at the scale $M_U$.
This potential serves a matching condition when we solve
one-loop renormalization group equations above and below
the scale $M_X$.
The potential is written down in terms of SUGRA, so it will be
useful to disclose the structure of SUGRA from the measurement
of SUSY spectrum.

We should consider the renormalization effects for the soft SUSY
breaking parameters and diagonalize the scalar mass matrix
$\langle V_{\hat{k}\hat{l}} \rangle$ to derive the weak
scale SUSY spectrum.

\section{Conclusion}

We have derived the low-energy effective theory starting from
non-minimal SUGRA with unified gauge symmetry under
some physical assumptions and requirements
in a model-independent manner.
The result is summarized in Eqs.(\ref{calWeff})--(\ref{<hatD>}).
We state chief results in correspondence with the assumptions.

The starting SUGRA consists of a non-minimal K\"alher potential
and a superpotential without the {\it hidden} ansatz
based on the hidden sector SUSY breaking scenario.
The non-minimality leads to non-universal soft SUSY breaking
terms as pointed out in Ref.\cite{SW}.
The dangerous $B$ term, which destabilizes the weak scale, can
exist if any conditions are not imposed
on Yukawa couplings in $W_{SG}$.

The SUGRA has a unified gauge symmetry which is broken down at a
scale $M_U$.
Some scalar fields get the VEVs of $O(M_U)$.
There exist heavy fields with the masses of $O(M_U)$ in addition to
light fields with the masses of $O(m_{3/2})$.
In such a situation, there appear extra non-universal contributions
to the soft SUSY breaking terms reflected to the combination of
the non-minimality of K\"alher potential and the breakdown of
extra gauge symmetry.
The most important one comes from
the $D$-term condensations of the heavy gauge sector.
This contribution is propotional to the charge of broken
diagonal generators, so we can know the large gauge symmetry
by the precision measurement of scalar masses.
There can exist many dangerous terms which threaten
to the gauge hierarchy, so we required that such terms
are suppressed.

It is expected that low-energy theories are checked by
the precision measurements of the SUSY spectrum
at the weak scale. From an optimistic point of view,
if the SUSY is realized
in nature, low-energy theories can be a touchstone
in elementary particle physics
at the beginning of the next century.

\section*{Acknowledgements}
The author is grateful to H.~Murayama, H.~Nakano and I.~Joichi
and especially M.~Yamaguchi for useful discussions.
This work is supported by the Grant-in-Aid for
Scientific Research ($\sharp$07740212) from
the Japanese Ministry of Education, Science and Culture.

\end{document}